\documentclass[aps,floatfix,twocolumn]{revtex4}
\usepackage{graphicx}
\usepackage{graphics,epsfig}
\usepackage{amssymb}
\def \P{{\cal P}}
\def \R{{\cal R}}
\begin{document}
\title{Constraining hybrid inflation models with WMAP three-year results}
\author{Antonio Cardoso\footnote{email: Antonio.Cardoso@port.ac.uk}}
\affiliation{Institute of Cosmology and Gravitation, University of
Portsmouth, Portsmouth, PO1 2EG, United Kingdom}
\date\today
\begin{abstract}
We reconsider the original model of quadratic hybrid inflation in
light of the WMAP three-year results and study the possibility of
obtaining a spectral index of primordial density perturbations,
$n_s$, smaller than one from this model. The original hybrid
inflation model naturally predicts $n_s\geq1$ in the false vacuum
dominated regime but it is also possible to have $n_s<1$ when the
quadratic term dominates. We therefore investigate whether there is
also an intermediate regime compatible with the latest constraints,
where the scalar field value during the last 50 e-folds of inflation
is less than the Planck scale.
\end{abstract}
\maketitle
\section{Introduction}
The results from the WMAP three-year data \cite{Spergel:2006hy} have
provided the first indication that the spectral index of primordial
density perturbations, $n_s$, is smaller than one. The original
hybrid inflation model \cite{Linde:1993cn} naturally predicts
$n_s\geq1$ in the false vacuum dominated regime but it is also
possible to have $n_s<1$ when the quadratic term dominates
\cite{Copeland:1994vg}. We are therefore interested in whether there
is also an intermediate regime compatible with the latest
constraints. Hybrid inflation models are attractive from a
theoretical point of view because the inflaton field $\phi$ is far
below the Plank scale during inflation, in the false vacuum
dominated regime, which makes these models easier to implement
within supergravity \cite{Lyth:1998xn}. Therefore we also
investigate whether there is an intermediate regime where the scalar
field value during the last 50 e-folds of inflation is less than the
Planck scale.

In this paper we study the original hybrid inflation model with a
quadratic potential for the inflaton. We want to analyze carefully
this model in light of the new WMAP results, exploring the space of
parameters of the model. We have run a code to calculate numerically
the spectral tilt of the scalar curvature perturbations, $n_s$, the
running of the spectral tilt, $\alpha_s$, and the tensor-to-scalar
ratio, $r$, assuming slow-roll, for a specific region of parameters
of the model. We study the cases for which we have $n_s<1$ and
compare the results obtained with the WMAP three-year results. We
use the $68\%$ and $95\%$ confidence level contours from WMAP only
and WMAP + SDSS taken from \cite{Kinney:2006qm}. Recently similar
works have used the WMAP three-year results to put constraints on
hybrid inflation models
\cite{deVega:2006hb,Peiris:2006ug,Martin:2006rs}.

We also study the inverted hybrid inflation model \cite{Lyth:1996kt}
with a quadratic potential for the inflaton (see Appendix).
\newpage
\section{The hybrid inflation model}
The potential for the hybrid inflation model is given by
\cite{Copeland:1994vg}
\begin{equation}\label{hybridpotential}
V(\phi,\chi)=\frac{1}{4}\lambda(\chi^2-\chi_0^2)^2+\frac{1}{2}m^2\phi^2+\frac{1}{2}\lambda^\prime\phi^2\chi^2,
\end{equation}
where $\phi$ is the inflaton, $\chi$ is called the ``waterfall"
field, $\lambda$ and $\lambda^\prime$ are coupling constants and
$\chi_0$ and $m$ are constant masses.

It is assumed that $\chi$ stays at the origin, which corresponds to
a false vacuum, while $\phi$ rolls down from a initially large
(positive) value until it reaches a critical value,
$\phi_c=\sqrt{\lambda/\lambda^\prime}M$, after which $\chi$ becomes
unstable (the effective mass-squared becomes negative) and rapidly
rolls down towards one of the true minima at $\chi=\pm \chi_0$. Then
$\phi$ goes to zero and starts to oscillate while $\chi$ will reach
the true minimum. Inflation will end either with the instability or
because of the end of slow-roll, as in the single field case,
depending on which occurs first.

Before $\phi$ reaches $\phi_c$ we can write the potential as a
function of $\phi$ only,
\begin{equation}\label{hybridpotentialchi0}
V(\phi)=M^4+\frac{1}{2}m^2\phi^2,
\end{equation}
where we wrote the false vacuum energy density $\frac{1}{4}\lambda
\chi_0^4$ as just $M^4$. We define the ratio between the false
vacuum energy density and the inflaton energy density as
\begin{equation}\label{energydensityratio}
E(\phi)=\frac{M^4}{\frac{1}{2}m^2\phi^2}.
\end{equation}
So if $E \gtrsim 1$ we have false vacuum dominated inflation and if
$E \ll 1$ we have almost chaotic inflation \cite{Linde:1983gd}.
\subsection{The dynamics}
The dynamics of hybrid inflation are given by the equation of motion
of the inflaton (we are assuming that the ``waterfall" field stays
at the origin, so it does not evolve) and the Friedmann equation,
\begin{eqnarray}\label{equationsdynamics1}
\ddot{\phi}+3H\dot{\phi}&=&-V^{\prime}(\phi),\\
H^2&=&\frac{8\pi}{3m_{Pl}^2}\left(\frac{1}{2}\dot{\phi}^2+V(\phi)\right),
\label{equationsdynamics2}
\end{eqnarray}
where $H=\dot{a}/a$ is the Hubble parameter, $a$ is the scale
factor, $m_{Pl}$ is the Planck mass (which we set equal to $1$), a
dot represents a derivative with respect to time and a prime denotes
a derivative with respect to the field $\phi$. We will use the
slow-roll approximation for our calculations, which is given by the
conditions
\begin{eqnarray}\label{slowrollparameters}
\epsilon(\phi)&\equiv&\frac{m_{Pl}^2}{16\pi}\left(\frac{V^{\prime}(\phi)}{V(\phi)}\right)^2\ll1,\\
\eta(\phi)&\equiv&\frac{m_{Pl}^2}{8\pi}\frac{V^{\prime\prime}(\phi)}{V(\phi)}\ll1.
\end{eqnarray}
With these approximations, Eqs. (\ref{equationsdynamics1}) and
(\ref{equationsdynamics2}) can be written as
\begin{eqnarray}\label{equationsdynamicsslowroll1}
3H\dot{\phi}&\simeq&-V^{\prime}(\phi),\\
H^2&\simeq&\frac{8\pi}{3m_{Pl}^2}V(\phi,\chi).
\label{equationsdynamicsslowroll2}
\end{eqnarray}
Then we can write, for the number of e-folds of expansion between
two field values $\phi_1$ and $\phi_2$,
\begin{equation}\label{numbere-folds}
N(\phi_1,\phi_2)\equiv
\ln\frac{a_2}{a_1}\simeq-\frac{8\pi}{m_{Pl}^2}\int^{\phi_2}_{\phi_1}\frac{V(\phi)}{V^{\prime}(\phi)}d\phi.
\end{equation}
For the potential in Eq. (\ref{hybridpotentialchi0}) we have
\begin{eqnarray}\label{slowrollparametersnumbere-folds1}
\eta(\phi)&=&\frac{m^2 m_{Pl}^2}{4\pi(2M^4+m^2\phi^2)},\\
\epsilon(\phi)&=&\frac{m^4\phi^2m_{Pl}^2}{4\pi(2M^4+m^2\phi^2)^2}=\frac{1}{2}\frac{8\pi}{m_{Pl}^2}\eta^2\phi^2,\label{slowrollparametersnumbere-folds2}\\
N(\phi_1,\phi_2)&\simeq&\frac{8\pi
M^4}{m^2m_{Pl}^2}\ln\frac{\phi_1}{\phi_2}+\frac{2\pi}{m_{Pl}^2}(\phi_1^2-\phi_2^2).
\label{slowrollparametersnumbere-folds3}
\end{eqnarray}

When the end of slow-roll occurs before the inflaton reaches
$\phi_{c}$ we identify the end of inflation with the condition
$\epsilon=1$, which occurs for the value of the inflaton field
\cite{Copeland:1994vg}
\begin{equation}\label{phiepsilon1}
\phi_{\epsilon}=\frac{m_{Pl}}{\sqrt{16\pi}}\left(1+\sqrt{1-\frac{32\pi}{m_{Pl}^2}\frac{M^4}{m^2}}\right).
\end{equation}
There is a second root at a smaller $\phi$, below which $\epsilon$
becomes smaller than unity again, but numerically it has been found
that slow-roll is not re-established before $\phi=0$
\cite{Copeland:1994vg}.

We see that if $32\pi M^4/m_{Pl}^2m^2>1$ then $\phi_{\epsilon}$ does
not exist at all, so in this case inflation has to end by
instability. If $32\pi M^4/m_{Pl}^2m^2\leqslant1$ it will end by
instability when $\phi=\phi_{c}$ if $\phi_{c}>\phi_{\epsilon}$ or by
the end of slow-roll when $\phi=\phi_\epsilon$ if
$\phi_{\epsilon}>\phi_{c}$.

We assume that cosmological scales exit the Hubble scale at least
$50$ e-folds before the end of inflation. In practice the actual
number of e-folds is dependent upon the details of reheating at the
end of inflation. Given that $\phi_c$ can be made arbitrarily small
by suitable choice of $\lambda/\lambda^\prime$ we leave $\phi_{50}$,
the value of $\phi$ when cosmological scales leave the Hubble scale,
as a free parameter to be determined by observations, subject only
to the restriction that we must have at least $50$ e-folds between
$\phi_{50}$ and $\phi_{\epsilon}$. If $\phi_{\epsilon}$ does not
exist then we can always have at least $50$ e-folds and any value of
$\phi_{50}$ is allowed.
\subsection{The perturbations}
The power spectrum of scalar curvature perturbations at horizon
crossing (when the comoving scale $k$ equals the Hubble radius,
$k=aH$, during inflation), which is conserved on large scales in
single field inflation, is given by, to leading order in the
slow-roll parameters, \cite{Bassett:2005xm}
\begin{equation}\label{amplitudeperturbation}
\P_{\R}(k)=\left.\left(\frac{H^2}{2\pi\dot{\phi}}\right)^2\right|_*,
\end{equation}
where the subscript $*$ indicates that the quantity is to be
evaluated at horizon crossing. By virtue of the slow-roll conditions
this formula gives a value of $\P_{\R}$ which is nearly independent
of $k$. For the potential in Eq. (\ref{hybridpotentialchi0}),
assuming slow-roll, Eq. (\ref{amplitudeperturbation}) can be written
as, using Eqs. (\ref{slowrollparameters}),
(\ref{equationsdynamicsslowroll1}),
(\ref{equationsdynamicsslowroll2}), (\ref{hybridpotentialchi0}) and
(\ref{slowrollparametersnumbere-folds2}),
\begin{equation}\label{amplitudeperturbationphi50}
\P_{\R}=\frac{16\pi}{3}\frac{(2M^4+m^2\phi_{50}^2)^3}{m_{Pl}^6m^4\phi_{50}^2}.
\end{equation}
So for each values of $M$ and $m$ we can find the values of
$\phi_{50}$ (there are several possible values for each combination
of $M$ and $m$) which satisfy the density perturbation amplitude,
for scales of cosmological interest, from the WMAP three-year
results \cite{Spergel:2006hy} (assuming that the running of the
spectral tilt is zero).

The spectral tilt for the scalar curvature perturbations, the
running of the spectral tilt and the tensor-to-scalar ratio can be
written as, to leading order in the slow-roll parameters,
\cite{Bassett:2005xm}
\begin{eqnarray}\label{spectraltiltrunningratioslowroll}
n_s&=&1-6\epsilon_*+2\eta_*,\\
\alpha_s&=&16\epsilon_*\eta_*-24\epsilon_*^2+2\xi_*^2,\\
r&=&16\epsilon_*,
\end{eqnarray}
where $\xi$ is a higher order slow-roll parameter and is equal to
zero for our specific potential, Eq. (\ref{hybridpotentialchi0}).
Then for this potential we can write
\begin{eqnarray}\label{spectraltiltslowroll}
n_s&=&1+\frac{m_{Pl}^2m^2(M^4-m^2\phi_{50}^2)}{\pi(2M^4+m^2\phi_{50}^2)^2},\\
\alpha_s&=&\frac{m_{Pl}^4m^6\phi_{50}^2(4M^4-m^2\phi_{50}^2)}{2\pi^2(2M^4+m^2\phi_{50}^2)^4},\\
r&=&\frac{4m_{Pl}^2m^4\phi_{50}^2}{\pi(2M^4+m^2\phi_{50}^2)^2}.
\end{eqnarray}
\subsection{Results and discussion}
We have run a code to calculate numerically $\phi_{50}$, $E$ and the
parameters $n_s$, $r$ and $\alpha_s$ at Hubble crossing (when
$\phi=\phi_{50}$), assuming slow-roll and using the definitions and
results from the previous sections, for $m$ between $10^{-4}$ and
$10^{-8}$ and $M$ between $10^{-2}$ and $10^{-5}$. For each values
of $m$ and $M$ we have selected the value of $\phi_{50}$ such that
the amplitude of the scalar curvature perturbations Eq.
(\ref{amplitudeperturbationphi50}) obeys the WMAP three-year results
\cite{Spergel:2006hy}. We discard the cases for which the number of
e-folds between $\phi_{50}$ and $\phi_{\epsilon}$ is smaller than
$50$ and those with a blue spectrum, i.e., we require $n_s<1$. The
energy density ratio, Eq. (\ref{energydensityratio}), is evaluated
at $\phi_{50}$.

\begin{figure}[htbp]
  \begin{center}
  \includegraphics[width=6.25cm]{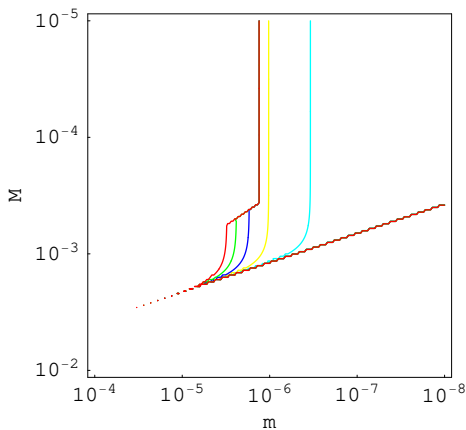}
  \end{center}
  \caption{Contour plot of $n_s$ as a function of $m$ and $M$ for $0.9 < n_s < 1$, for the hybrid inflation model. The red, green, blue, yellow and light blue contours (from left to right) represent, respectively, $n_s$ equal to $0.91$, $0.93$, $0.95$, $0.97$ and $0.99$.}
  \label{LCTPns50}
\end{figure}

We first note that as one goes from larger to smaller values of $M$
the density of points selected by the code gets larger because the
selection of the values of $M$ is logarithmic, as one can see
looking to the plot in Figure \ref{LCTPns50}.

\begin{figure}[htbp]
  \begin{center}
  \includegraphics[width=8cm]{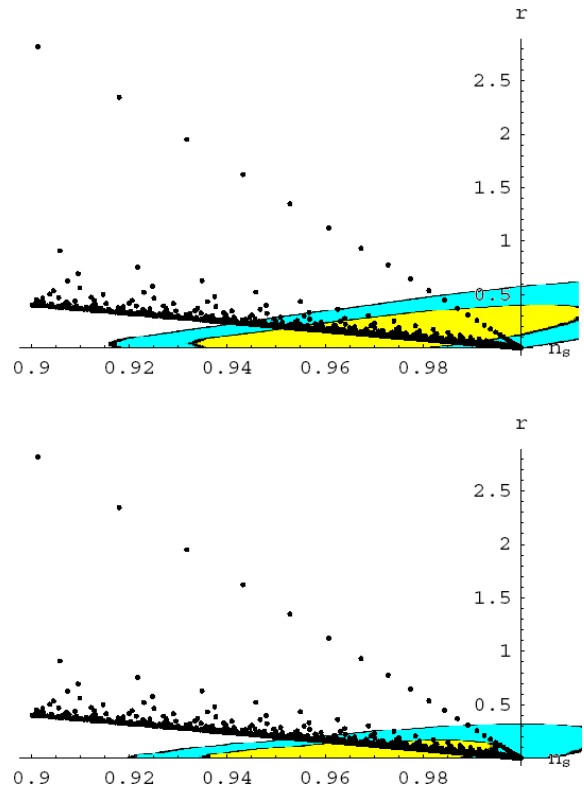}
  \end{center}
  \caption{Plots of $r$ as a function of $n_s$ for $0.9 < n_s < 1$ (the black dots represent the different values of $M$ and $m$), for the hybrid inflation model, with the $68\%$ (yellow) and $95\%$
(blue) confidence level contours from WMAP (top) and WMAP + SDSS
(bottom), taken from \cite{Kinney:2006qm}.}
  \label{LPTr50ns50}
\end{figure}

Analyzing the top plot in Fig. \ref{LPTr50ns50} we see that there is
a small range of parameters for which $(n_s,r)$ is inside the $68\%$
confidence level contour from WMAP only \cite{Kinney:2006qm}. For
this range we find
\begin{equation}\label{nsrwmapagreement68}
0.9525 \lesssim n_s \lesssim 0.9975~~~~and~~~~0.05 \lesssim r
\lesssim 0.35.
\end{equation}
For the range of parameters for which $(n_s,r)$ is inside the $95\%$
confidence level contour we find
\begin{equation}\label{nsrwmapagreement95}
0.94 \lesssim n_s \lesssim 1~~~~and~~~~0 \lesssim r \lesssim 0.55.
\end{equation}
Considering the results from WMAP + SDSS \cite{Kinney:2006qm} we see
that, looking to the bottom plot in Fig. \ref{LPTr50ns50}, the range
of parameters for which $(n_s,r)$ is inside the $68\%$ confidence
level contour is smaller than in the previous case. For this range
we find
\begin{equation}\label{nsrwmapsdssagreement68}
0.96 \lesssim n_s \lesssim 0.995~~~~and~~~~0.05 \lesssim r \lesssim
0.175.
\end{equation}
For the range of parameters for which $(n_s,r)$ is inside the $95\%$
confidence level contour we have
\begin{equation}\label{nsrwmapsdssagreement95}
0.95 \lesssim n_s \lesssim 1~~~~and~~~~0 \lesssim r \lesssim 0.3.
\end{equation}

\begin{figure}[htbp]
  \begin{center}
  \includegraphics[width=8cm]{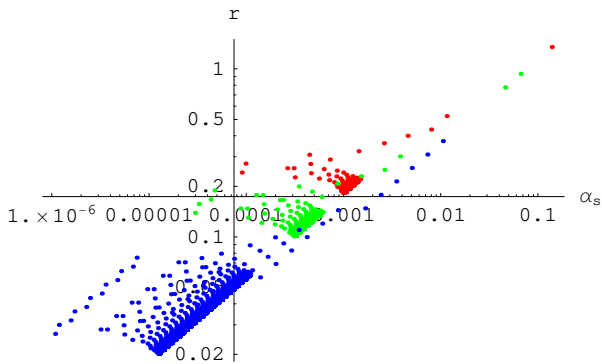}
  \end{center}
  \caption{Plot of $r$ as a function of $\alpha_s$ for $0.945<n_s<0.955$ (red), $0.965<n_s<0.975$ (green) and
$0.985<n_s<0.995$ (blue), for the hybrid inflation model.}
  \label{LogLPTr50alpha50095_7_9}
\end{figure}

\begin{figure}[htbp]
  \begin{center}
  \includegraphics[width=8cm]{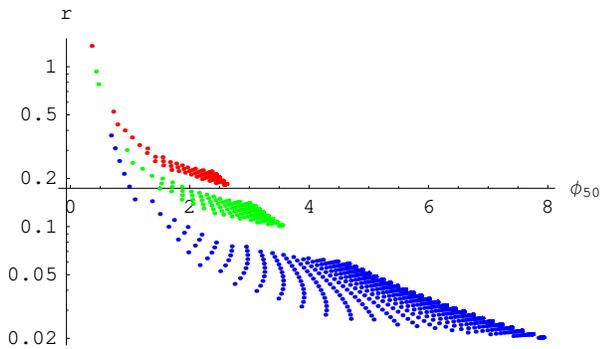}
  \end{center}
  \caption{Plot of $r$ as a function of $\phi_{50}$ for $0.945<n_s<0.955$ (red),
$0.965<n_s<0.975$ (green) and $0.985<n_s<0.995$ (blue), for the
hybrid inflation model.}
  \label{LogLPTr50phi50095_7_9}
\end{figure}

\begin{figure}[htbp]
  \begin{center}
  \includegraphics[width=8cm]{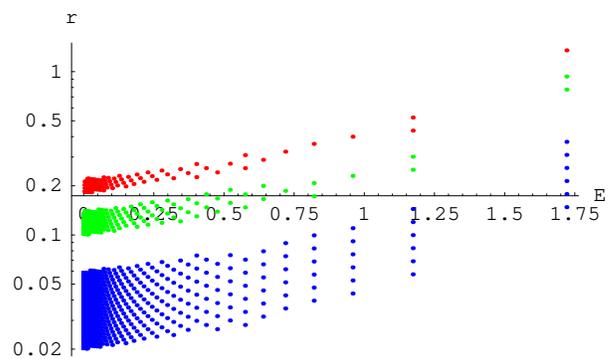}
  \end{center}
  \caption{Plot of $r$ as a function of $E$ for $0.945<n_s<0.955$ (red), $0.965<n_s<0.975$ (green) and
$0.985<n_s<0.995$ (blue), for the hybrid inflation model.}
  \label{LogLPTr50DR50095_7_9}
\end{figure}

Observing the plot in Figure \ref{LogLPTr50alpha50095_7_9} we see
that the closer one gets to large values of $M$ the larger are the
values of $r$ (i.e., the energy scale of inflation increases when
$M$ increases) and $\alpha_s$. Analyzing the plot in Figure
\ref{LogLPTr50phi50095_7_9} we see that the closer one gets to small
values of $\phi_{50}$ the larger are the values of $r$ (and so also
the values of $\alpha_s$ get larger). We also note that $\phi_{50}$
is never much smaller than $1$, i.e., the Planck mass. Looking to
the plot in Figure \ref{LogLPTr50DR50095_7_9} we note that the
energy density ratio is never much larger than $1$, which means that
there is never a real false vacuum domination of the energy density,
otherwise we would have $n_s\geq1$. We also see that the closer one
gets to the false vacuum domination cases (which occur for large
values of $M$) the larger are the values of $r$ and the larger (and
positive) are the values of the the running $\alpha_{50}$.

From this analysis we see that when we have false vacuum domination
and small values of $\phi_{50}$, i.e., when we are more distant from
chaotic inflation, then the values of the the running $\alpha_{50}$
are large and positive, which is in contradiction with the WMAP
three-year results \cite{Kinney:2006qm}. Also the values of the
tensor-to-scalar ratio $r$ tend to be very large for these cases,
which is also in disagreement with the observations.

When $M$ goes to very small values we recover the results for
chaotic inflation, which correspond to the lower limit for $r$ along
the values of $n_s$ in Fig. \ref{LPTr50ns50}. For these cases, for
which $\phi_{50}$ is very large and the energy density ratio very
small, we have small values for $\alpha_{50}$ and $r$, which is in
very good agreement with the WMAP three-year results
\cite{Kinney:2006qm}.
\section{Conclusions}
We have found that there is an intermediate regime for the original
hybrid inflation model compatible with $n_s<1$, but as one
approaches the false vacuum dominated limit within this regime the
tensor-to-scalar ratio, $r$, and the running of the spectral tilt,
$\alpha_s$, become large (with a positive running), which is in
contradiction with the WMAP three-year results \cite{Kinney:2006qm}.
This agrees with the results in
\cite{deVega:2006hb,Peiris:2006ug,Martin:2006rs}. We found a lower
bound on the allowed values of $r$, which might be an observational
signal for hybrid inflation because if there is an upper bound on
the spectral index, $n_s<1$, then we find a lower bound on the
tensor-to-scalar ratio. This lower bound corresponds to chaotic
inflation, which is in good agreement with the new WMAP results
\cite{Kinney:2006qm}. We also saw that it is difficult to get
$\phi_{50}$ smaller than the Planck scale because decreasing
$\phi_{50}$ requires us to increase the vacuum energy scale, $M$,
and this increases the tensor-to-scalar ratio, $r$.

The inverted hybrid inflation model (see Appendix) is in much better
agreement with the WMAP three-year results \cite{Kinney:2006qm} than
the original hybrid inflation model when we consider the false
vacuum dominated regimes, especially because there is no lower bound
on the allowed values of $r$. Moreover, in the inverted hybrid model
there is no problem obtaining small values for $\phi_{50}\ll 1$,
which makes these model easier to implement within supergravity
\cite{Lyth:1998xn}.
\section*{Acknowledgements}
I would like to thank David Wands for very useful discussions and
comments on this work. I would also like to thank the authors of
Ref. \cite{Kinney:2006qm} for permission to reproduce likelihood
contours from their work. I am supported by FCT (Portugal) PhD
fellowship SFRH/BD/19853/2004.
\appendix
\section*{Appendix: The inverted hybrid inflation model}
If we invert the sign of the inflaton energy density in Eq.
(\ref{hybridpotentialchi0}) we get
\begin{equation}\label{smallfieldpotential}
V(\phi)=M^4-\frac{1}{2}m^2\phi^2,
\end{equation}
which corresponds to the potential for the inverted hybrid inflation
model with a quadratic potential for the inflaton. A particular case
of this model are the small-field inflation models
\cite{Bassett:2005xm}, for which the potential in Eq.
(\ref{smallfieldpotential}) corresponds to a Taylor expansion about
the origin and higher order terms are required to provide a
potential minimum at some $\phi\neq 0$, as required to connect to a
reheating stage (which is necessary to make the model realistic). So
the inflaton starts with an initial small (positive) value and rolls
down towards the minimum at a larger value, then it starts to
oscillate and inflation ends with the end of slow-roll.

All the equations obtained before for the original hybrid inflation
model, except Eq. (\ref{phiepsilon1}), are still valid but with
$m^2$ replaced by $-m^2$. To find the value of $\phi_{\epsilon}$ we
search for the solutions of Eq.
(\ref{slowrollparametersnumbere-folds2}) with $\epsilon$ equal to
$1$ (there are two different solutions for this equation but the
results do not depend on which solution we choose).
\subsection{Results and discussion}
In this case we have run the same code (with the appropriate changes
in the equations) but for $m$ between $10^{-4}$ and $10^{-10}$ and
$M$ between $10^{-2}$ and $10^{-6}$. The energy density ratio, Eq.
(\ref{energydensityratio}), is evaluated $50$ e-folds after the
inflaton has the value $\phi_{50}$.

\begin{figure}[htbp]
  \begin{center}
  \includegraphics[width=6.25cm]{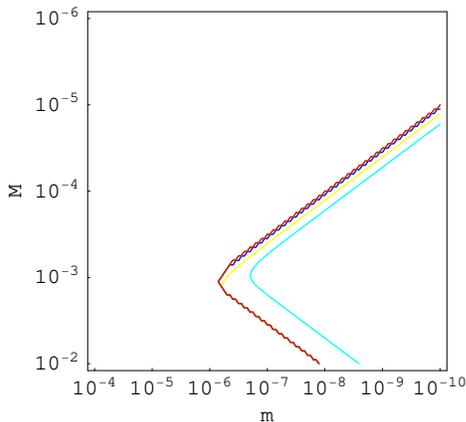}
  \end{center}
  \caption{Contour plot of $n_s$ as a function of $m$ and $M$ for $0.9 < n_s < 1$, for the inverted hybrid inflation model. The red, green, blue, yellow and light blue contours (from left to right) represent, respectively, $n_s$ equal to $0.91$, $0.93$, $0.95$, $0.97$ and $0.99$.}
  \label{LCPTns50new}
\end{figure}

Note that also in this case the selection of the values of $M$ is
logarithmic, as one can see looking to the plot in Figure
\ref{LCPTns50new}.

\begin{figure}[htbp]
  \begin{center}
  \includegraphics[width=8cm]{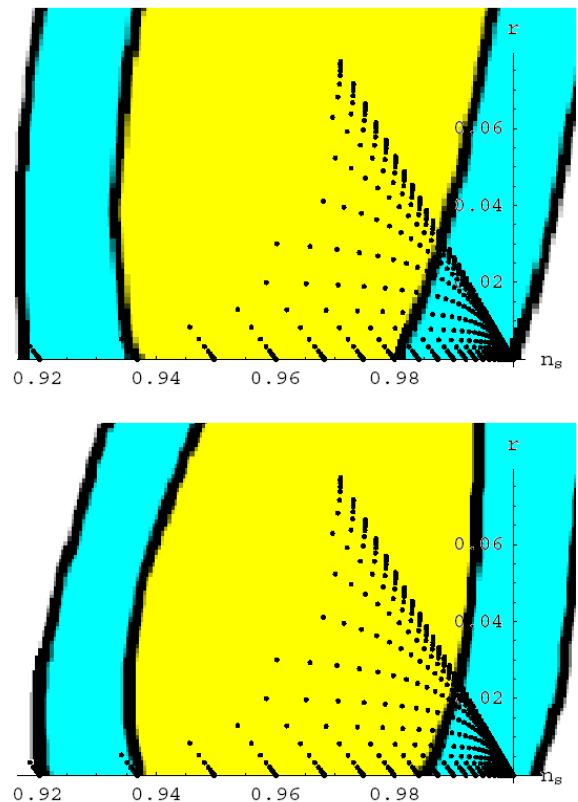}
  \end{center}
  \caption{Plots of $r$ as a function of $n_s$ for $0.9 < n_s < 1$ (the black dots represent the different values of $M$ and $m$), for
the inverted hybrid inflation model, with the $68\%$ (yellow) and
$95\%$ (blue) confidence level contours from WMAP (top) and WMAP +
SDSS (bottom), taken from \cite{Kinney:2006qm}.}
  \label{LPTr50ns50new}
\end{figure}

Analyzing the top plot in Fig. \ref{LPTr50ns50new} we see that there
is a much wider range of parameters for which $(n_s,r)$ is inside
the $68\%$ confidence level contour from WMAP only
\cite{Kinney:2006qm} than in the original hybrid case, especially
because there is no lower bound on the allowed values of $r$, in
contrast with the original hybrid inflation model. For this range we
find
\begin{equation}\label{nsrinvertedwmapagreement68}
0.935 \lesssim n_s \lesssim 0.9875~~~~and~~~~0 \lesssim r \lesssim
0.0775.
\end{equation}
For the range of parameters for which $(n_s,r)$ is inside the $95\%$
confidence level contour we get
\begin{equation}\label{nsrinvertedwmapagreement95}
0.9175 \lesssim n_s \lesssim 1~~~~and~~~~0 \lesssim r \lesssim
0.0775.
\end{equation}
Observing the bottom plot in Fig. \ref{LPTr50ns50new} we can see
that there is also a wide range of parameters for which $(n_s,r)$ is
inside the $68\%$ confidence level contour from WMAP + SDSS
\cite{Kinney:2006qm}. For this range we get
\begin{equation}\label{nsrinvertedwmapsdssagreement68}
0.935 \lesssim n_s \lesssim 0.9925~~~~and~~~~0 \lesssim r \lesssim
0.0775,
\end{equation}
and for the range of parameters for which $(n_s,r)$ is inside the
$95\%$ confidence level contour we find
\begin{equation}\label{nsrinvertedwmapsdssagreement95}
0.92 \lesssim n_s \lesssim 1~~~~and~~~~0 \lesssim r \lesssim 0.0775.
\end{equation}

\begin{figure}[htbp]
  \begin{center}
  \includegraphics[width=8cm]{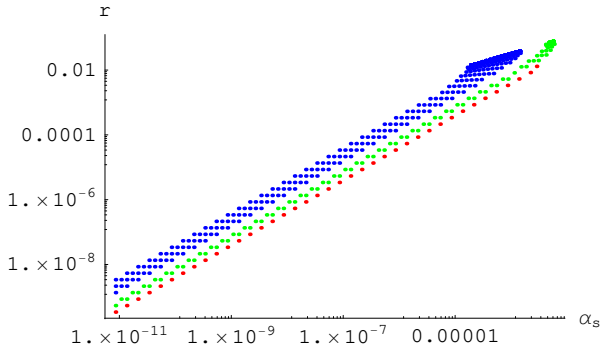}
  \end{center}
  \caption{Plot of $r$ as a function of $\alpha_s$ for $0.945<n_s<0.955$ (red), $0.965<n_s<0.975$ (green) and
$0.985<n_s<0.995$ (blue), for the inverted hybrid inflation model.}
  \label{LogLPTr50alpha50095_7_9new}
\end{figure}

\begin{figure}[htbp]
  \begin{center}
  \includegraphics[width=8cm]{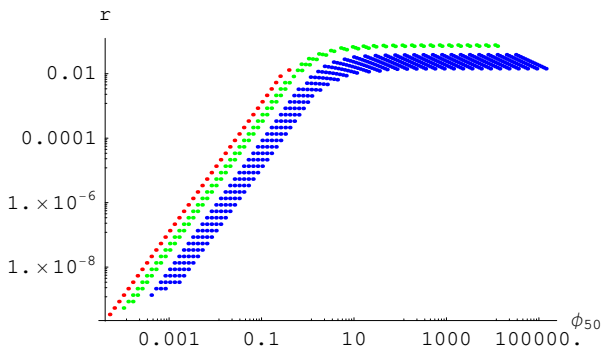}
  \end{center}
  \caption{Plot of $r$ as a function of $\phi_{50}$ for $0.945<n_s<0.955$ (red),
$0.965<n_s<0.975$ (green) and $0.985<n_s<0.995$ (blue), for the
inverted hybrid inflation model.}
  \label{LogLPTr50phi50095_7_9new}
\end{figure}

\begin{figure}[htbp]
  \begin{center}
  \includegraphics[width=8cm]{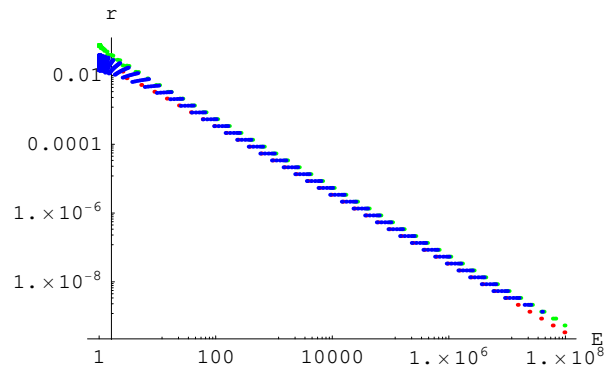}
  \end{center}
  \caption{Plot of $r$ as a function of $E$ for $0.945<n_s<0.955$ (red), $0.965<n_s<0.975$ (green) and
$0.985<n_s<0.995$ (blue), for the inverted hybrid inflation model.}
  \label{LogLPTr50DR50095_7_9new}
\end{figure}

Looking to the plot in Figure \ref{LogLPTr50alpha50095_7_9new} we
see that the closer one gets to large values of $M$ the smaller are
the values of $r$ and $\alpha_s$ (although we note that they are
always small, even in the cases for which $M$ is small). Analyzing
the plot in Figure \ref{LogLPTr50phi50095_7_9new} we see that the
closer one gets to small values of $\phi_{50}$ the smaller are the
values of $r$ (and so also the values of $\alpha_s$ get smaller). We
note that here $\phi_{50}$ can be much smaller than the Planck mass.
Observing the plot in Figure \ref{LogLPTr50DR50095_7_9new} we see
that the closer one gets to the more false vacuum domination cases
(which occur for large values of $M$) the smaller are the values of
$r$ (and so also the values of $\alpha_s$). In this case we note
that the energy density ratio can be very large, so we can have a
strong vacuum domination.

From the previous analysis we see that for the inverted hybrid
inflation model we can have, specially for large $M$, very small
values for $r$ and $\alpha_s$, which is in very good agreement with
the WMAP three-year results \cite{Kinney:2006qm}. For these cases
$\phi_{50}$ is also very small (it can be much smaller than $1$,
i.e., the Planck mass) and $E$ is much larger than $1$, which means
that there is a large false vacuum domination of the energy density.
Therefore we can conclude that this model is in much better
agreement with the WMAP results than the original hybrid inflation
model when we consider the false vacuum domination regimes,
especially because there is no lower bound on the allowed values of
$r$. Moreover, in this model there is no problem obtaining small
values for $\phi_{50}\ll 1$, which makes these model easier to
implement within supergravity \cite{Lyth:1998xn}.
\newpage
{}
\end{document}